\documentclass[final,english]{bullsrsl}

\usepackage[latin1]{inputenc}

\usepackage[T1]{fontenc}

\usepackage{natbib} 
\usepackage{graphicx}

\begin{document}
\title{Unveiling photometric transients with the ILMT and their follow-up using DOT}

\author[affil={1}, corresponding]{Priyanshi}{Kumari}
\author[affil={2,1}]{Bhavya}{Ailawadhi}
\author[affil={3,4}]{Talat}{Akhunov}
\author[affil={1,5}]{Monalisa}{Dubey}
\author[affil={1,5}]{Naveen}{Dukiya}
\author[affil={6}]{Sara}{Filali}
\author[affil={7,8}]{Paul}{Hickson}
\author[affil={1}]{Gokul Singh}{Mehra}
\author[affil={1}]{Kuntal}{Misra}
\author[affil={9}]{Vibhore}{Negi}
\author[affil={1,10}]{Kumar}{Pranshu}
\author[affil={6,1}]{Jean}{Surdej}
\author[affil={1,10}]{Sarvesh Kumar}{Yadav}

\affiliation[1]{Aryabhatta Research Institute of Observational Sciences (ARIES), Manora Peak, Nainital-263002, India}

\affiliation[2]{Physical Research Laboratory, Navrangpura, Ahmedabad, Gujarat-380009, India}

\affiliation[3]{National University of Uzbekistan, Department of Astronomy and Astrophysics, 100174 Tashkent, Uzbekistan}

\affiliation[4]{Ulugh Beg Astronomical Institute of the Uzbek Academy of Sciences, Astronomicheskaya 33, 100052 Tashkent, Uzbekistan}

\affiliation[5]{Mahatma Jyotiba Phule Rohilkhand University, Pilibhit Bypass Road, Bareilly 243006, Uttar Pradesh, India
}
\affiliation[6]{Institute of Astrophysics and Geophysics, Li\`ege University, All\'ee du 6 Ao\^ut 19c, 4000 Li\`ege, Belgium
}
\affiliation[7]{Department of Physics and Astronomy, The University of British Columbia, 6224 Agricultural Road, Vancouver, BC V6T 1Z1, Canada
}

\affiliation[8]{Outer Space Institute, The University of British Columbia, 325-6224 Agricultural Road, Vancouver, BC V6T 1Z1, Canada
}

\affiliation[9]{Kavli Institute for Astronomy and Astrophysics, Peking University, Beijing 100871, China}

\affiliation[10]{Department of Applied Optics and Photonics, University of Calcutta, Kolkata, 700106, India}

\correspondance{kumaripriyanshi98@gmail.com}

\maketitle

\begin{abstract}
The two 4-m class telescopes at the Devasthal site in India are complementary: the 4-m International Liquid Mirror Telescope (ILMT) is dedicated to the discovery of new transient phenomena, which can be followed up using the 3.6-m Devasthal Optical Telescope (DOT) for their optical characterisation. The ILMT continuously surveys a $22.3^\prime \times 22.3^\prime$ wide strip of the zenith sky. A dedicated transient detection and classification pipeline, \texttt{PyLMT}, based on optimised image subtraction and convolutional neural network (CNN) techniques, enables the discovery of numerous photometric transients with the ILMT. However, these transient discoveries still require spectroscopic confirmation to determine their types, for which spectra obtained with the DOT are essential. Here, we present a selection of photometric transients, specifically supernova candidates. We investigate their photometric and spectroscopic properties using ILMT light curves, supplemented with ZTF data and DOT spectroscopic observations.

\end{abstract}

\keywords{ILMT, transients, follow-up, light curves, spectroscopy}

\section{Introduction}
The International Liquid Mirror Telescope (ILMT) is a 4-m, zenith-pointing optical telescope that uses rotating liquid mercury as its primary mirror. It is installed at the Devasthal Observatory (ARIES), located at an altitude of 2378 m in the Nainital district of Uttarakhand, India \citep{2000A&AS..144..349S}. The telescope has an effective focal length of 9.44 m and operates in the Time-Delay Integration (TDI) mode, in which the CCD charge is shifted synchronously with the motion of celestial sources across the focal plane due to Earth's rotation. Because the telescope remains fixed at the zenith and stellar images follow slightly curved trajectories on the detector, a dedicated five-element optical corrector is employed to minimize optical distortions \citep{10.1093/mnras/258.3.543, Hickson_1998}. Nightly observations are obtained with a 4k $\times$ 4k CCD camera positioned at the prime focus, using the Sloan $g'$, $r'$, and $i'$ filters. The details of the telescope can be found in \cite{2025A&A...694A..80S}. Owing to its fixed zenith-pointing configuration, the telescope scans the same strip of sky every clear night, making it ideally suited for wide-field time-domain surveys with a primary focus on detecting transient and variable astrophysical phenomena, followed by spectroscopic characterisation with the 3.6m Devasthal Optical Telescope (DOT). This observing strategy enables the discovery and long-term monitoring of transient and variable sources, including supernovae (SNe), cataclysmic variables (CVs), active galactic nuclei (AGN), and other variable stars. The repeated observations also facilitate the construction of deep photometric catalogs and support studies of gravitational lensed candidates and other rare astrophysical phenomena. The scientific capabilities of the ILMT have been demonstrated through a variety of studies, including the detection and characterization of SNe \citep{2024BSRSL..93..880K}, Solar System objects \citep{2024BSRSL..93..941P, 2026arXiv260702675F}, low-surface-brightness galaxies \citep{2024BSRSL..93..918F}, multiply imaged quasars \citep{2024BSRSL..93..933A} and space debris \citep{2024BSRSL..93..903H}. In this work, we present three spectroscopically confirmed SN candidates \textit{viz.} SN~2023af (Section~\ref{sec:2023af}), 2024cjb (Section~\ref{sec:2024cjb}), and 2025dov (Section~\ref{sec:2025dov}). These SNe were detected with the ILMT and subsequently followed up with the 3.6m DOT. 
A brief summary of this work is presented in Section~\ref{sec:summary}.

\section{SN 2023af: The first SN in the ILMT field}
\label{sec:2023af}
The ILMT achieved its first light on 29$^{\mathrm{th}}$ April 2022. The first SN in the ILMT field of view was identified during commissioning, when the \texttt{PyLMT} pipeline \citep{2025MNRAS.538..133P} was still under development. SN~2023af was discovered by cross-matching the ILMT field with the online database of SNe reported in 2023. As a result, SN~2023af was identified in the host galaxy MCG+05-20-043, making it the first SN identified in the ILMT field of view. The SN was independently discovered by the XOSS collaboration on 2023 January~2.9~UT at R.A.\,$=11^{\rm h}04^{\rm m}36^{\rm s}$ and Dec.\,$=+29^\circ31^\prime01^{\prime\prime}$ (J2000.0). The discovery magnitude was reported to be $16.73 \pm 0.10$\,mag in an unfiltered (clear) band. A spectrum obtained using the Yunnan Faint Object Spectrograph and Camera (YFOSC) mounted on the Lijiang 2.4\,m telescope revealed prominent hydrogen features characteristic of a young Type~II SN. Following its detection with ILMT, low-resolution spectroscopic observations were triggered using the 3.6m DOT under the Target of Opportunity (ToO) program.  Follow-up spectroscopic observations were performed at two epochs on 2023 March 13 and March 26, corresponding to approximately 71 and 84 days after discovery, respectively. Spectroscopic analysis confirmed that SN~2023af is a Type IIP SN. The photometric and spectroscopic analysis of SN~2023af is presented in \cite{2024BSRSL..93..880K}.

\section{SN 2024cjb}
\label{sec:2024cjb}

\subsection{Detection and Observations}

SN~2024cjb was first identified by the ILMT on 14$^{\mathrm{th}}$ February 2024 at 18:06~UT using the automated \texttt{PyLMT} transient detection pipeline, at RA = $09^{\mathrm h}\,11^{\mathrm m}\,27.47^{\mathrm s}$ and Dec = $+29^\circ\,29'\,36.55''$. The source showed a clearly associated host galaxy and was therefore considered a promising SN candidate. A simple cone-search with the Transient Name Server (TNS) indicated that the source had already been reported by the ALeRCE broker \citep{2021AJ....161..242F} for the Zwicky Transient facility \citep[ZTF;][]{2019PASP..131a8002B}. The original discovery by ALeRCE/ZTF \citep{2024cjb_2024TNSTR.405....1F} occurred two days prior to our detection, on 12$^{\mathrm{th}}$ February 2024 at 06:52~UT. The last non-detection reported in the ZTF was on 2024 January 31 in the $g^\prime$-band, with a limiting magnitude of 19.66 mag. However, using ILMT data, the last non-detection epoch could be refined to 11$^{\mathrm{th}}$ February 2024 at 18:17~UT in $r^\prime$-band with a limiting magnitude of approximately 21.7 mag. The host galaxy of the SN was identified as the catalogued galaxy SDSS~J091127.17+292934.7. 

Spectroscopic follow-up observations of SN~2024cjb were obtained with the ARIES Devasthal Faint Object Spectrograph and Camera \citep[ADFOSC;][]{Omar_2019} mounted on DOT on 7$^{\mathrm{th}}$ and 8$^{\mathrm{th}}$ March 2024, 24 and 25 days after the discovery, respectively. An earlier spectrum was acquired by the ZTF using the Palomar 60-inch telescope on 5$^{\mathrm{th}}$ March 2024. 

\begin{figure}[htbp]
    \centering
    \includegraphics[width=0.7\linewidth]{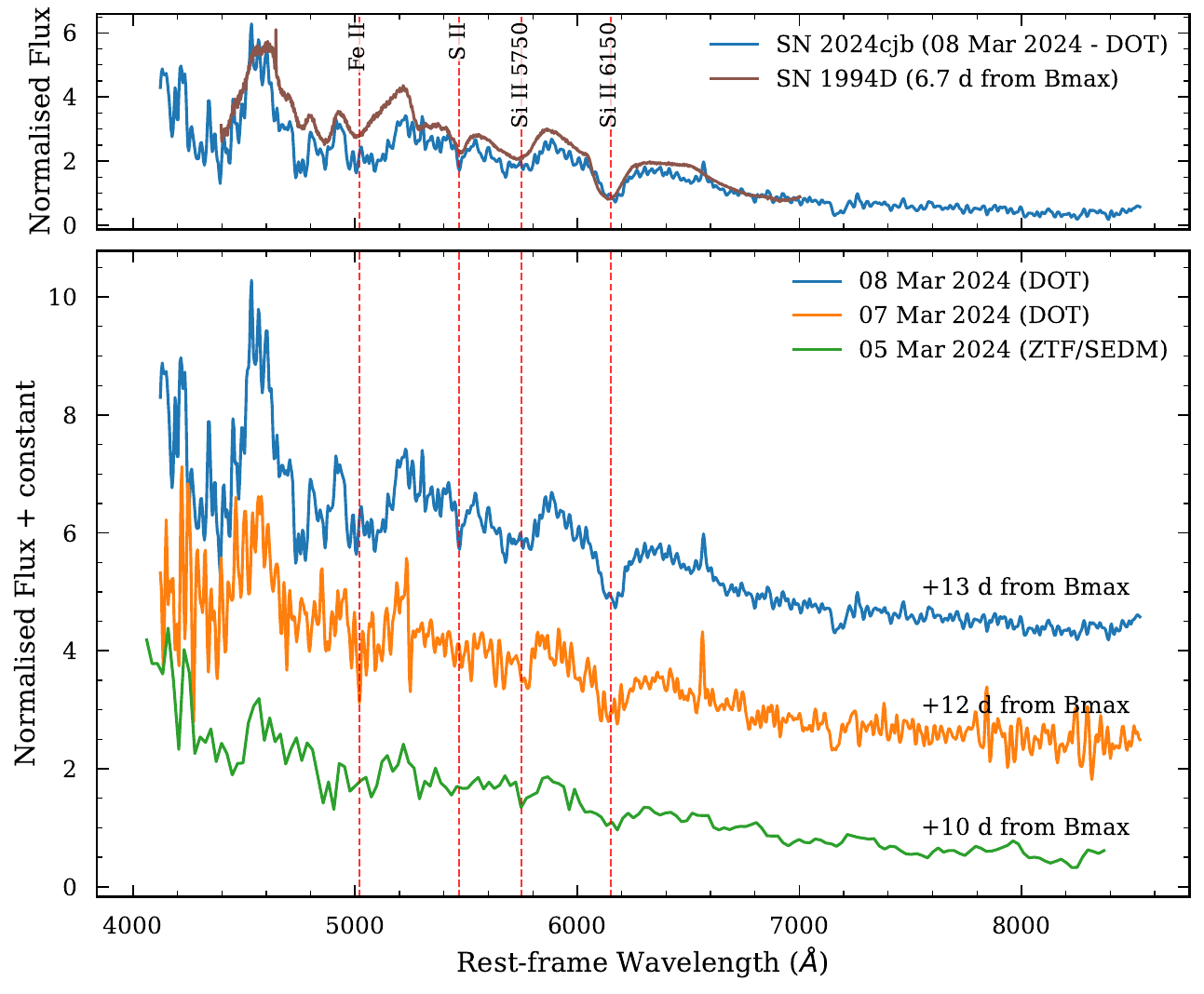}
    \caption{Low-resolution optical spectra obtained with the 3.6\,m DOT using ADFOSC on 2024 March 7 and 8, along with the Palomar 60-inch spectrum from 2024 March 5 retrieved from WISeREP, are plotted in the lower panel. The best-match \texttt{GELATO} spectrum is shown in the upper panel for comparison.}
    \label{fig:cjb_combined}
\end{figure}

\begin{figure}[t]
    \centering
    \includegraphics[width=0.48\textwidth]{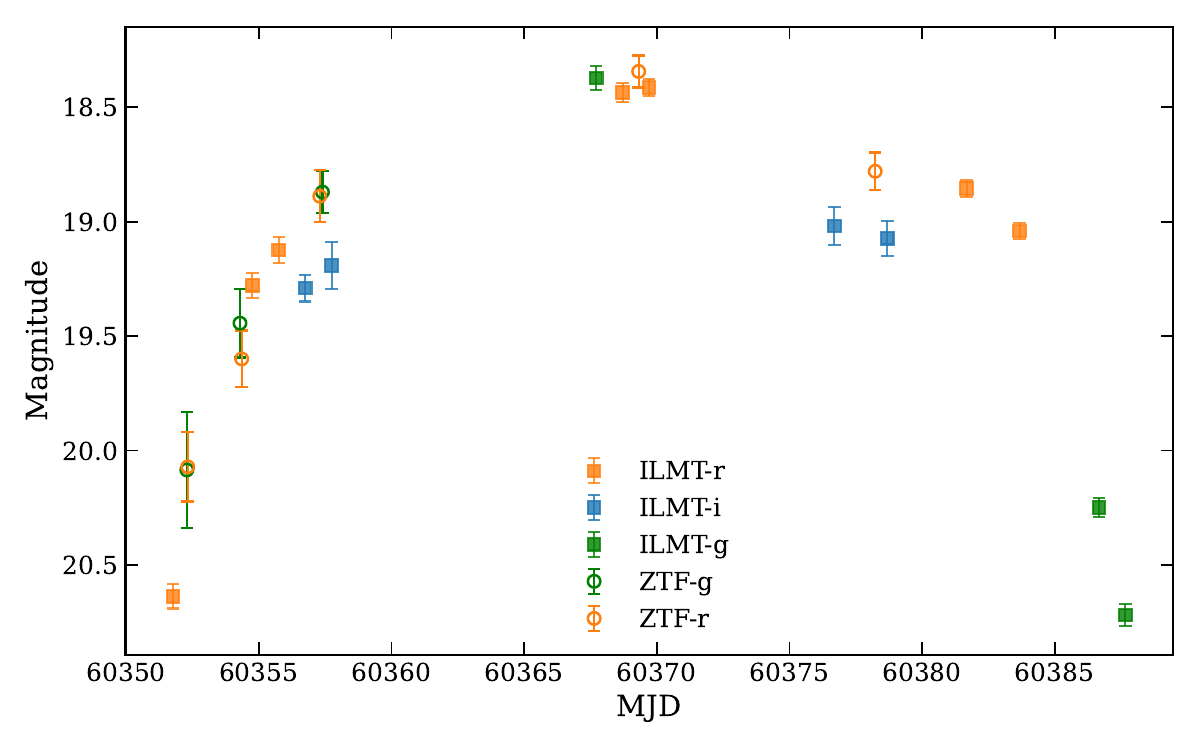}
    \hfill
    \includegraphics[width=0.48\textwidth]{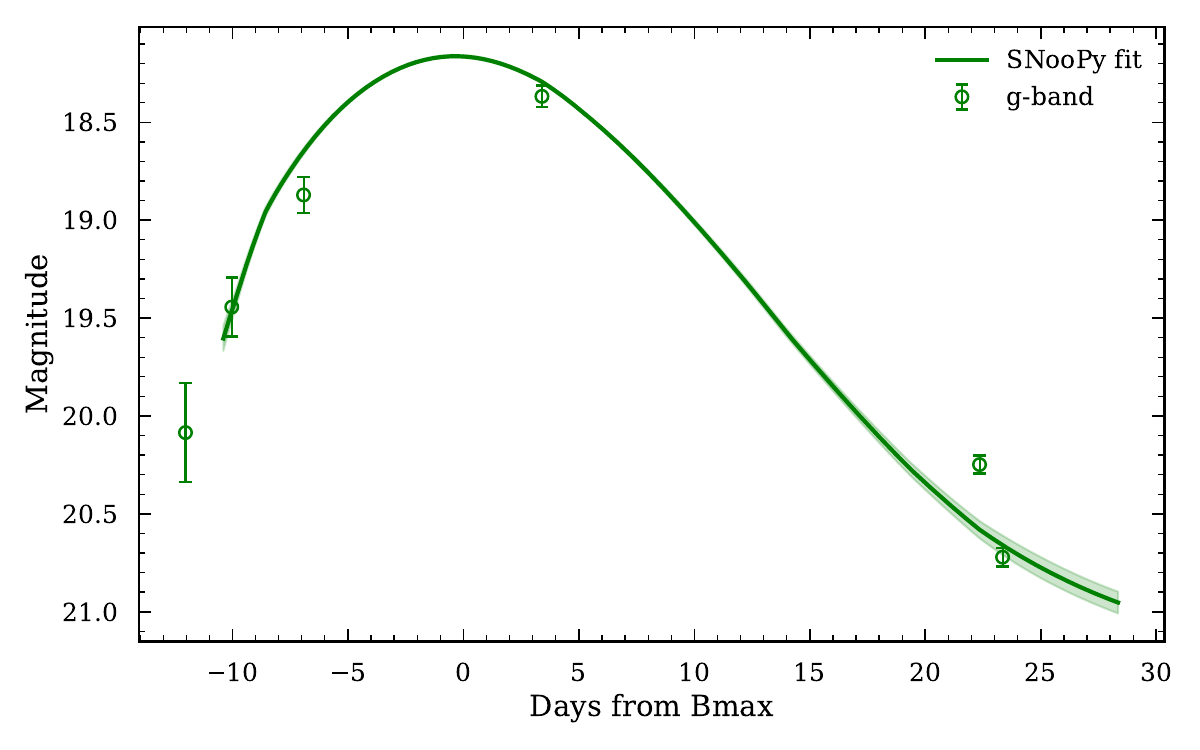}

    \includegraphics[width=0.48\textwidth]{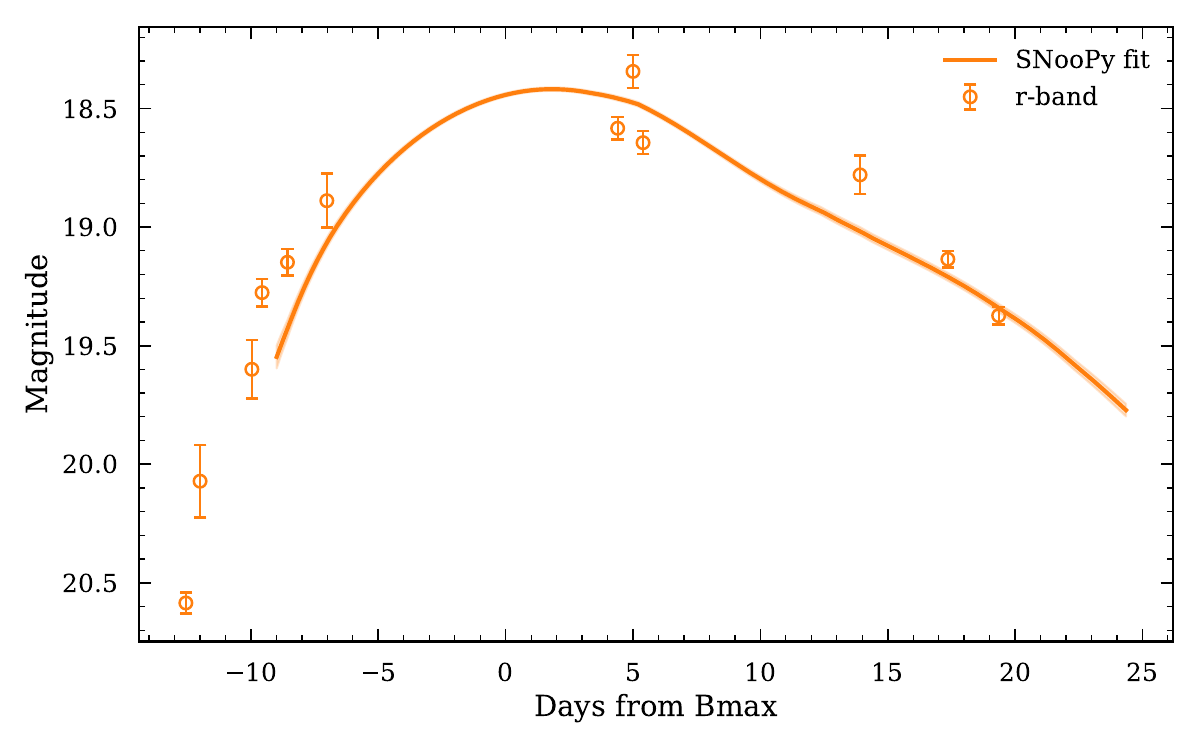}
    \hfill
    \includegraphics[width=0.48\textwidth]{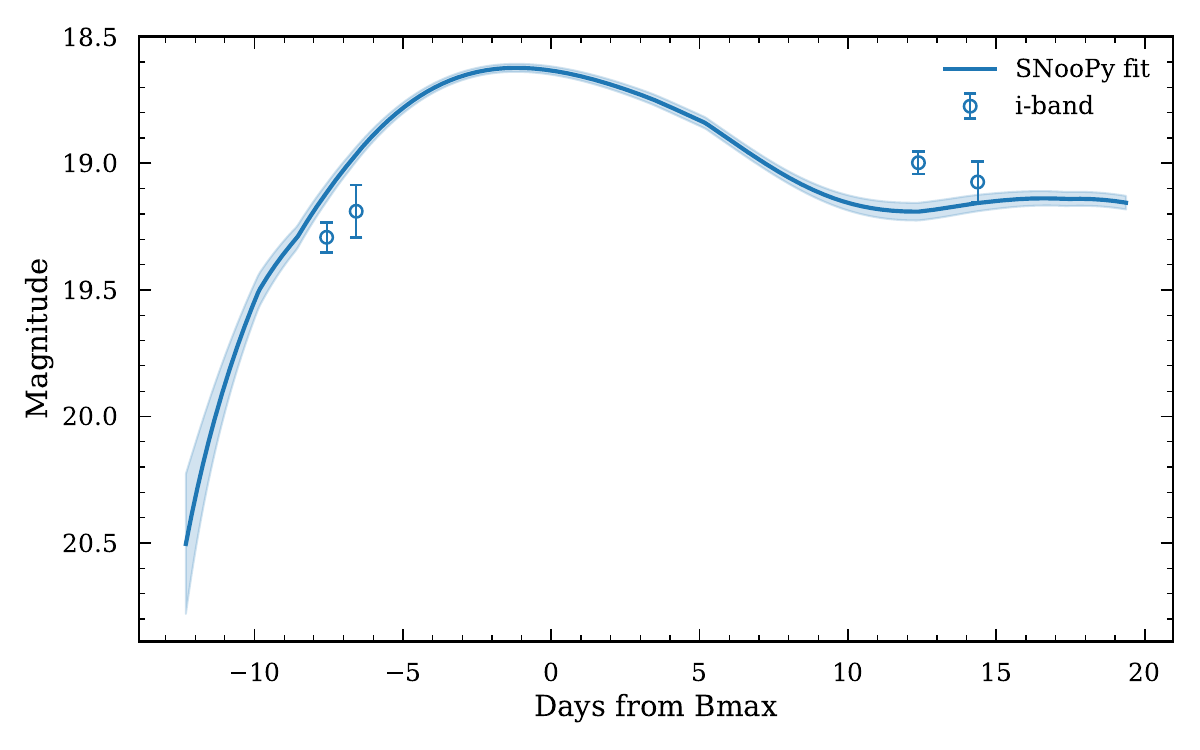}

    \caption{Observed $g'$, $r'$, and $i'$ light curves of SN~2024cjb (upper left) and the corresponding \texttt{SNooPy} \texttt{max\_model} fits to the $g'$, $r'$, and $i'$ bands (remaining panels). The solid curves show the best-fitting templates, while the shaded regions denote the $1\sigma$ template uncertainties returned by \texttt{SNooPy}.}
    \label{fig:sn2024cjb_snpy}
\end{figure}

\subsection{Spectroscopic and Photometric study of SN~2024cjb}

The spectra (shown in Figure \ref{fig:cjb_combined}) of exhibited a prominent Si\,II absorption feature near 6150~\AA,~which occurs in the case of Type~Ia SN \citep{2024TNSCR.611....1J}. Such SNe arise from the thermonuclear disruption of carbon--oxygen white dwarfs in binary systems and, owing to their relatively uniform peak luminosities, serve as standardizable candles for measuring extragalactic distances and constraining cosmological parameters. Furthermore, we compared the spectrum acquired on 2024 March 8 with SN spectral templates using the \textsc{GELATO} \citep{2008A&A...488..383H} classification tool (shown in the top panel of Figure \ref{fig:cjb_combined}). The best match was obtained with SN~1994D at approximately +6.7 days relative to $B$-band maximum, consistent with a normal Type~Ia SN.

Post-detection, the source was detected with the ILMT in the SDSS g$'$, r$'$ and i$'$--bands across multiple epochs. The ILMT photometry was supplemented with ZTF photometric data to improve phase coverage. The top left panel in Figure~\ref{fig:sn2024cjb_snpy} shows the multi-band light curve of SN~2024cjb. To further characterize the photometric evolution of SN~2024cjb, we fit the $g'$-, $r'$ and $i'$-band light curves using the SuperNovae in Object-Oriented Python \citep[\texttt{SNooPy};][]{2011AJ....141...19B} package (see Figure~\ref{fig:sn2024cjb_snpy}). We adopted the \texttt{max\_model} parameterized by the stretch parameter ($st$), which simultaneously fits the multi-band light curves to determine the epoch of $B$-band maximum and the peak magnitudes. The best-fitting model yields a stretch parameter of $st = 0.768 \pm 0.041$ and an epoch of $B$-band maximum at $\mathrm{MJD}=60364.28 \pm 0.38$. The fitted peak apparent magnitudes are $g'_{\rm max}=18.30 \pm 0.09$ mag at $\mathrm{MJD}=60363.96 \pm 0.38$ \ and $r'_{\rm max}=18.47 \pm 0.04$ mag at $\mathrm{MJD}=60365.64 \pm 0.39$. Relative to the first detection, the rise times to maximum are $12.20 \pm 0.38$ days in the $g'$-band and $13.88 \pm 0.38$ days in the $r'$ band. The corresponding 15-day decline parameters are $\Delta m_{15}(g')=1.28 \pm 0.08$ mag and $\Delta m_{15}(r')=0.68 \pm 0.02$ mag. The slower decline in the $r'$ band compared to the $g'$ band is consistent with the expected photometric evolution of normal Type~Ia SNe. Detailed analysis could not be performed for $i'$-band due to insufficient data. The best-fitting light curve models are shown in Figure~\ref{fig:sn2024cjb_snpy}.

\section{SN 2025dov}
\label{sec:2025dov}

\subsection{Detection and Observations}

SN~2025dov was detected by the ILMT on 19$^{\mathrm{th}}$ March 2025 at 
RA = $11^{\mathrm h}\,16^{\mathrm m}\,00.00^{\mathrm s}$ and Dec = $+29^\circ\,21'\,48.00''$, with an apparent detection magnitude of 18.04 in the $r'$ filter. It was detected by the ILMT 13 days after its original discovery by the ATLAS group \citep{2018PASP..130f4505T}, which reported a magnitude of 19.2 in the orange ATLAS filter, detected with the ATLAS Mauna Loa telescope. The event was classified as a Type Ia SN by the Spectroscopic Classification of Astronomical Transients (SCAT) Survey, using the Supernova Integrated Field Spectrograph (SNIFS) mounted on the UH88 telescope on 8\textsuperscript{th} March 2025. Spectroscopic follow-up of the transient source was also performed with the DOT (ADFOSC) on 27$^{\mathrm{th}}$ March 2025.

\begin{figure}[t]
    \centering
    \includegraphics[width=0.7\textwidth]{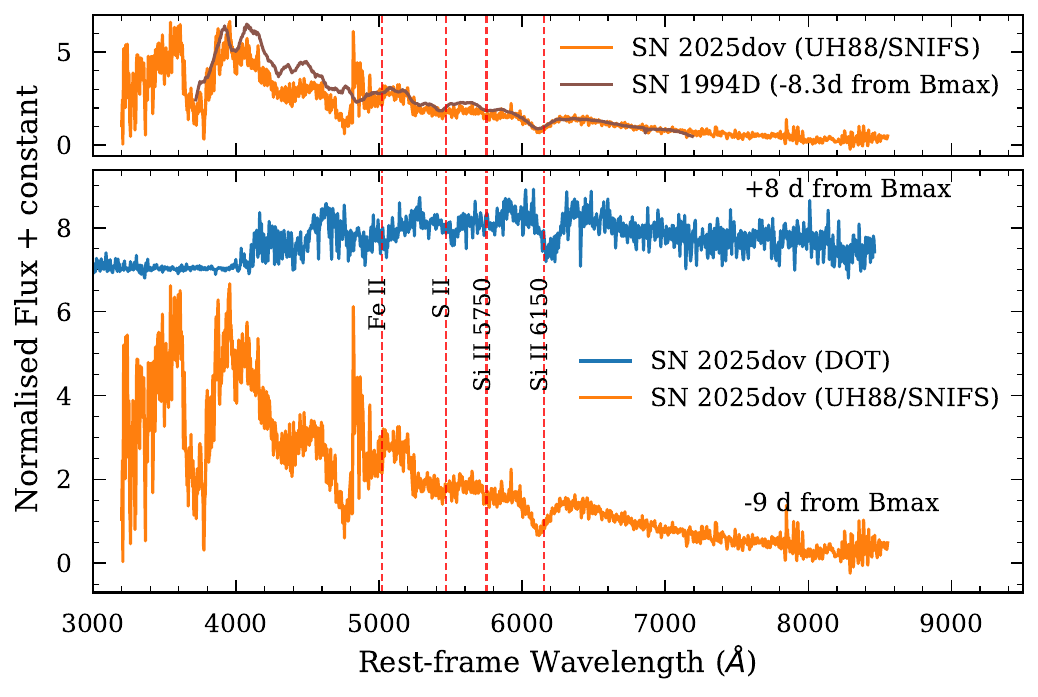}

    \vspace{-2mm}
    \caption{Spectroscopic evolution of SN~2025dov. The upper panel compares the classification spectrum with the best-matching \textsc{GELATO} template, while the lower panel shows the spectral evolution over the available epochs in the rest frame.}
    \label{fig:SN2025dov_spec}
    \vspace{-2mm}
\end{figure}

\begin{figure}[t]
    \centering
    \includegraphics[width=0.49\textwidth]{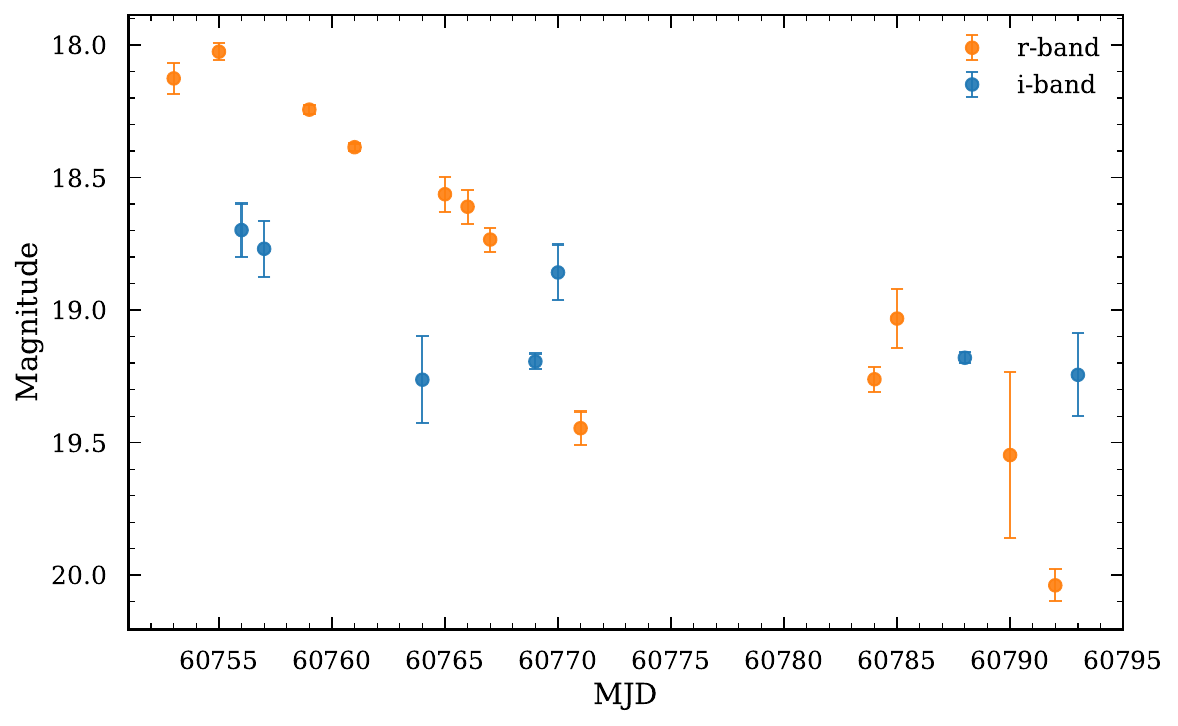}
    \hfill
    \includegraphics[width=0.49\textwidth]{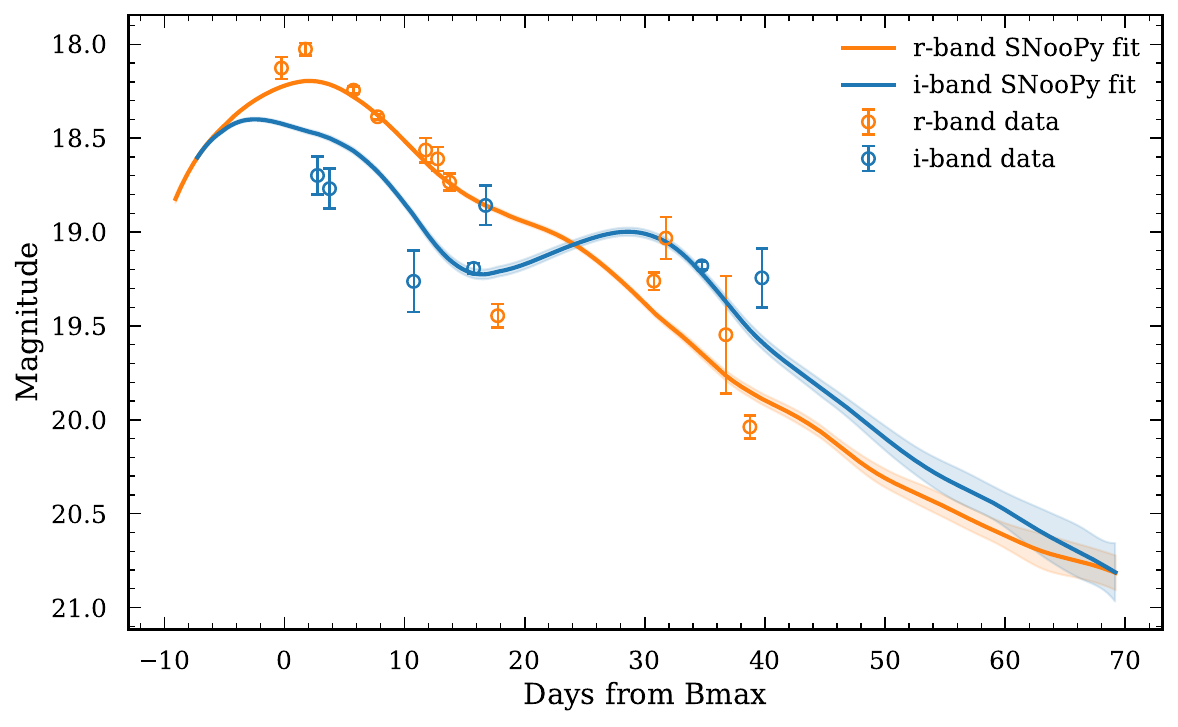}

    \vspace{-2mm}
    \caption{Observed $r'$- and $i'$-band light curves of SN~2025dov (left) and the corresponding \texttt{SNooPy} \texttt{max\_model} fit (right). The solid curves represent the best-fitting templates, while the shaded regions indicate the $1\sigma$ template uncertainties returned by \texttt{SNooPy}.}
    \label{fig:SN2025dov_phot}
    \vspace{-2mm}
\end{figure}

\subsection{Spectroscopic and Photometric study of SN~2025dov}


Figure~\ref{fig:SN2025dov_spec} shows the ADFOSC spectrum along with the classification spectrum from the SNIFS instrument. The Si II absorption feature at 6150~\AA~ can be clearly identified in both spectra, consistent with a Type~Ia SN. The SNIFS spectrum was also classified through template matching with the GELATO software. The best agreement was obtained with the Type~Ia SN, SN 1994D at an epoch of $-8.3$ d with respect to B-band maximum, as presented in the top panel of Figure~\ref{fig:SN2025dov_spec}.

The optical light curves of SN~2025dov in the $r'$ and $i'$ bands were fitted using the \texttt{\texttt{SNooPy} max\_model} to characterize their photometric evolution (see Figure~\ref{fig:SN2025dov_phot}). The best-fitting model yields a stretch parameter of $st = 0.992 \pm 0.103$ and an epoch of $B$-band maximum at $\mathrm{MJD}=60753.26 \pm 1.89$. The fitted peak magnitude in the $r'$ band is $18.20 \pm 0.08$ mag, occurring at $\mathrm{MJD}=60755.37 \pm 1.48$, corresponding to only $2.37 \pm 1.48$ days from the first ILMT detection. The $r'$-band light curve declines by $\Delta m_{15}(r)=0.659 \pm 0.043$ mag over the first 15 days after maximum. For the $i'$ band, the model-inferred peak magnitude is $18.40 \pm 0.11$ mag at $\mathrm{MJD}=60750.79 \pm 2.07$.

\section{Summary}
\label{sec:summary}

The 4-m ILMT has played a vital role in detecting and discovering new photometric transients, primarily SNe, especially after its integration with the \texttt{PyLMT} pipeline. So far, a total of 22 SN candidates have been detected, although many of them still lack spectroscopic confirmation. In this work, we present the detection and follow-up of three spectroscopically confirmed SNe (2023af, 2024cjb, and 2025dov). SN~2023af was classified as a Type IIP SN, whereas both SNe~2024cjb and 2025dov were classified as Type Ia SNe based on the spectroscopic observations acquired with the 3.6m DOT and other publicly available spectroscopic data. 

With the ILMT efficiently surveying the zenith sky across multiple bands, many more such photometric transients are expected to be detected in the future. Multi-band light curves for an SN candidate are crucial for studying various photometric parameters, such as rise time, peak magnitude, decline rate, and $^{56}$Ni mass. However, photometric detection alone is insufficient and should be supplemented with prompt spectroscopic follow-up with the 3.6m DOT for detailed progenitor characterisation and more accurate determination of the explosion mechanism. Such a synergistic operation of the 4m ILMT and the 3.6m DOT enables a comprehensive and coordinated transient science program.

\begin{acknowledgments}
 The 4-m International Liquid Mirror Telescope (ILMT) project results from a collaboration between the Institute of Astrophysics and Geophysics (University of Li\`{e}ge, Belgium), the Universities of British Columbia, Laval, Montreal, Toronto, Victoria, and York University, and Aryabhatta Research Institute of Observational Sciences (ARIES, India). The team acknowledges the contributions of ARIES's technical and scientific staff. JS wishes to thank Service Public Wallonie, F.R.S.-FNRS (Belgium), and the University of Li\`{e}ge, Belgium, for funding the construction of the ILMT. PH acknowledges financial support from the Natural Sciences and Engineering Research Council of Canada, RGPIN-2019-04369. PH, SF, and JS thank ARIES for hospitality during their visits to Devasthal. We sincerely acknowledge Anna Pospiezalska for her significant contributions to ILMT functioning. MD acknowledges the Innovation in Science Pursuit for Inspired Research (INSPIRE) fellowship award (DST/INSPIRE Fellowship/2020/IF200251) for this work. KP acknowledges the support from the Erasmus+ Programme of the European Union for a research visit to the Institute of Astrophysics and Geophysics, University of Li\`ege, Belgium. TA thanks the Ministry of Higher Education, Science, and Innovations of Uzbekistan (grant FZ-20200929344). JS and KM acknowledge the assistance received from the ANRF SERB-726 VAJRA Faculty Scheme, India. KM acknowledges the support from the BRICS grant DST/ICD/BRICS/Call-5/CoNMuTraMO/2023 (G) funded by the DST, India.
\end{acknowledgments}

\begin{furtherinformation}
\begin{orcids}
\end{orcids}

\begin{authorcontributions}
This work results from a long-term collaboration to which all authors have made significant
contributions.
\end{authorcontributions}

\begin{conflictsofinterest}
The authors declare no conflict of interest.
\end{conflictsofinterest}

\end{furtherinformation}

\bibliographystyle{bullsrsl-en}
\bibliography{BINA_Priyanshi}

\end{document}